\documentclass{article}
\usepackage{makeidx}  
\usepackage[utf8]{inputenc}
\usepackage[T1]{fontenc} 
\usepackage{amssymb}
\usepackage{graphicx}
\usepackage{booktabs}
\usepackage{todonotes}
\usepackage{caption}
\usepackage{stmaryrd}
\usepackage{url}
\usepackage[pdfpagelabels,hypertexnames=false,breaklinks=true,bookmarksopen=true,bookmarksopenlevel=2]{hyperref}
\newtheorem{example}{Example}
\begin{document}

%
\title{A Cross-Layer Security Analysis for Process-Aware Information Systems}
%

%

 \author{Maria Leitner$^1$, Zhendong Ma$^1$, Stefanie Rinderle-Ma$^2$\\
 $^1$Austrian Institute of Technology, \\Digital Safety \& Security Department\\ 
  $^2$University of Vienna, Faculty of Computer Science,  \\Research Group Workflow Systems and Technology\\
\{maria.leitner, zhendong.ma\}@ait.ac.at\\
 stefanie.rinderle-ma@univie.ac.at}

\date{}
\maketitle              

\begin{abstract}
Information security in Process-aware Information System (PAIS) relies on many factors, including security of business process and the underlying system and technologies. Moreover, humans can be the weakest link that creates pathway to vulnerabilities, or the worst enemy that compromises a well-defended system. Since a system is as secure as its weakest link, information security can only be achieved in PAIS if all factors are secure. 
In this paper, we address two research questions: how to conduct a cross-layer security analysis that couple security concerns at business process layer as well as at the technical layer; and how to include human factor into the security analysis for the identification of human-oriented vulnerabilities and threats. We propose a methodology that supports the tracking of security interdependencies between functional, technical, and human aspects which contribute to establish a holistic approach to information security in PAIS. We demonstrate the applicability with a scenario from the payment card industry.

\end{abstract}

\section{Introduction}
\label{sec:Introduction}

Humans are the weakest link when it comes to security in everyday business processes (e.g., \cite{mitnick_art_2001} and \cite{sasse_transforming_2001}). For example, the 2014 data breach report of \cite{verizon_2014_2014} shows that as of 2013, more than 85\% of data breaches stem from external agents, more than 10\% from insiders and around 1\% from business partners. 
Research shows that for example insiders (i.e., employees) intentionally circumvent security measures for three reasons (\cite{kirlappos_learning_2014}): First, employees \textsl{lack security-awareness} and have no incentive to apply security-compliant behavior. Secondly, employees skip security procedures in order to fulfill their job effectively. The cost of adherence to compliance is too high. Lastly, compliance is \textsl{impracticable} and employees choose different ways to fulfill their jobs. However, this non-compliant behavior increases the likelihood for vulnerabilities because it is hard to control (see \cite{briney_security_2000} and \cite{walton_balancing_2006}). While business still keeps running, such behavior is also difficult to detect until an incident occurs.  

Business process can be seen as a vehicle to integrate the human, functional, and technical perspective in an enterprise. As humans are an essential part for fulfilling and managing business processes \cite{kabicher-fuchs_human-centric_2012}, it is important to incorporate them when it comes to security in business processes.
Although security in Process-aware Information Systems (PAIS) is a interdisciplinary research field that faces many challenges such as human orientation or an agreement of technology, it often centers on the specification of access control mechanisms (cf. \cite{leitner_systematic_2014}). However, these mechanisms are often neither evaluated with users nor directly visible to the user and find ways to bypass it (e.g., \cite{briney_security_2000} and \cite{bartsch_how_2013}).
Most research that incorporates human involvement focuses on the design and specification of security requirements in process modeling languages via extensions. 
For example, modeling extensions for UML or BPMN (e.g., \cite{wolter_modelling_2008} and \cite{rodriguez_bpmn_2007}) exist that represent specific restrictions such as privacy, integrity, or confidentiality requirements on tasks. However, the concrete implementation on the technical level or user studies that evaluate the e.g., comprehensibility or cognitive effectiveness are often missing. 

In this paper we aim to address two main research questions: (a) how to conduct a cross-layer security analysis that couples security concerns at business process layer as well as at the technical layer; and (b) how to include human factor into the security analysis for the identification of human-oriented vulnerabilities and threats. 
With question (a), we target to identify security concerns that \textsl{go beyond the business process} and to investigate technical aspects and human aspects that contribute to a thorough and systematic security analysis. Question (b) aims to provide a \textsl{holistic view} that shows that human involvement is as important as technical aspects when it comes to securing business processes. 

In order to address questions (a) and (b), security is investigated as a cross-layer concern in PAIS architectures and a multilayer PAIS architecture model is provided that incorporates humans as essential factor spanning across the business process, system design, and implementation layers. 
The multilayer PAIS model is used as an input to generate a method for a human-oriented security analysis that contains four steps: business process analysis, function mapping, technical analysis, and human factor analysis. Moreover, we analyze and evaluate our method with a scenario from the payment card industry. With this examination, we show that humans are a key factor in the design, development, and execution of secure business processes. The approach follows the conceptual framework set out in \cite{dhillon2001current}. 

The method proposed in this paper contributes to security engineering in PAIS but also to other research disciplines (e.g., human-centric PAIS \cite{kabicher-fuchs_human-centric_2012}). We envision that it can be used for systematic, risk-driven security analysis of PAIS and guiding the design and development of the system. 

The paper is structured as follows:
Section~\ref{sec:SecurityInBusinessProcesses} analyzes security in business processes based on a multilayer model that incorporates technical and human aspects. 
Section~\ref{sec:Analysis} describes the method for the human-oriented security analysis with an application scenario from the payment card industry. 
The main findings, limitations, and impact on research and practice are outlined in Section~\ref{sec:Discussion}. Section~\ref{sec:RelatedWork} describes related work and Section~\ref{sec:Conclusion} concludes the paper.

  \section{Security in PAIS Architectures}
\label{sec:SecurityInBusinessProcesses}

Security in business processes is an interdisciplinary field that links security mechanisms from the business process management and security research community (cf.~\cite{leitner_systematic_2014}). When considering security from a holistic, engineering point of view, we have to assess not only the system and the architecture, but also the human-related aspects of the business process. To establish a structured view of the PAIS architecture, three layers that are relevant to security analysis can be identified (e.g., \cite{bohr2013business}):

\begin{itemize}
	\item the \textbf{business process layer} represents the functional organization, 
	\item the \textbf{system design layer} centers on the specification of the system architecture and its components, 
	\item and the \textbf{implementation layer} incorporates the technologies utilized to implement the system.
\end{itemize}
In addition, a fourth \textbf{human layer} is considered orthogonally to the other three layers. In the following, we will discuss each layer in more detail. 


\subsection{Business Process Layer}
\label{sec:BusinessProcessLayer}
The business process layer centers on the processes when organizing and managing work in an organization (cf. \cite{dumas_fundamentals_2013}), i.e. the functional organization. This includes not only the process participants but also the outcomes and the customer values of processes. Business processes can be found in every business domain such as in banks, hospitals, or retail stores. Attacking a business process requires full or at least partial knowledge of the \textsl{participants} in a process and on the \textsl{activities} performed. 

\subsection{System Design Layer}
\label{sec:SystemDesignLayer}

System design defines system architecture, system components and their interfaces, and data. Putting together, they realize the specification of business processes. System design can be done at different levels of abstraction, from conceptualization and modeling of systems to selection of concrete technologies and system artifacts. System design follows engineering principles which usually includes analysis, design, and verification.

Generally, a system architect needs to consider common ICT components such as hardware, software, database and networks. Within the system architecture, software components such as applications and services perform various actions defined by a business process. During the design, the input/output to the software components and their behaviors are considered to a great extent to ensure that all components will behave according to the design. Also considered are the component-component and component-user interactions.     


With respect to information security, all of the variants at the system design layer can have profound security implications on business process security, as well as on system implementation. To achieve security by design, a set of security controls need to be analyzed and planned for securing business processes (cf. \cite{leitner_systematic_2014}). Besides, security testing can be carried out at design time (cf. \cite{mouratidis_security_2007}). 

\subsection{Implementation Layer}
\label{sec:ImplementationLayer}
The implementation layer is concerned with the technologies utilized to implement the system design and covers an extensive set of technologies (e.g., web services, programming languages), PAIS products (e.g., SAP, ERP, or Oracle), and protocols (e.g., transport protocols, networks) which can be utilized to implement and execute business processes. Hence, the underlying technology is a key factor when it comes to security in PAIS. For example, a web service-based application system can be attacked with e.g., denial of service, XML, or SOAP attacks (cf. \cite{jensen_soa_2007,vorobiev_security_2006}).
With the increase in use of different technologies and system complexity, the number of vulnerabilities increases and therefore the attack surface and attack vector. 

Depending on the technology used, different outsiders might be more interested or motivated to attack. For instance, a web-based application might be an easier target for non-professional (e.g., script kiddies) and professional hackers than a closed system (which might be only interesting for hired professionals). 

\subsection{Human Layer}
\label{sec:HumanLayer}

In business processes, manual (i.e. performed by a human) and automated tasks (i.e. performed by a machine or program) are distinguished. 
Typically, humans access PAIS based on their roles that represent job functions or organizational affiliations (\cite{fuchs_roles_2011}). 
To clearly identify human involvement and its impact on security, we analyze human activities based typical roles they take in business processes (see \cite{dumas_fundamentals_2013,weske_business_2007,kavantzas_web_2005,omg_business_2011}). Typically, managers (i.e. \textsl{chief process officer}, \textsl{business engineers}, and \textsl{process owners}) use dashboards or control panels to evaluate and track business processes (such as handover time between tasks). \textsl{Process designers}, \textsl{system architects}, and \textsl{developers} are responsible for the design and implementation of business process models and systems and the set up of security controls. 
\textsl{Process participants} and \textsl{knowledge workers} actively participate in business processes. 
 \textsl{Subjects} participate passively in the business process (e.g., as a patient in a medical procedure). 
\textsl{Business partners} take part in a business process choreography where public views on the internal processes are visible to the partners. Each business partner has to comply with a defined set of rules or a set of contractual agreements (see \cite{weske_business_2007}). This implicates also security-related policies (\cite{carminati_web_2005,hafner_seaas-reference_2009}.







\textsl{Employee, Business partner, Contractor, Nation state, Professional hacker, Non-professional hacker}
is a list of attacker roles adapted from commonly used attacker types in security literature and practice, and the threat agents library defined in~\cite{Casey:2010:TAN:1852666.1852728}. From an information security point of view, humans are the main actor that passively or actively influences the security of a system. Generally, humans have two types of ``negative'' influences on security: they either introduce vulnerabilities in terms of flaws or mistakes in the design, implementation, configuration, and operation of the system (\cite{DBLP:journals/compsec/KraemerCC09}), or pose threats as attackers to exploit the vulnerabilities and comprise the security of the system (\cite{mitnick_art_2001,okenyi_anatomy_2007}). Obviously, different roles typically connected to a business process are able to introduce vulnerabilities and launch attacks. 
For example, social engineering attacks are well known for acquiring passwords or other information from humans in order to gain access to information systems (\cite{granger_social_2001}). Hence, common tasks in risk assessment are to analyze and estimate how likely a part of the system could be attacked, by whom, and with what means. Although actual attacks are often done by crafted input, computer instructions or programs (e.g. malware), the origins of the attacks can always be linked to humans. Therefore, identifying potential attackers and their roles is crucial for security analysis of PAIS.

  \section{Cross-Layer Security Analysis}
\label{sec:Analysis}

This section proposes a methodology for cross-layer security analysis of business processes including a human-oriented perspective. The methodology is based on the PAIS architecture (cf. Section \ref{sec:SecurityInBusinessProcesses}) and divided into four steps. Each step analyzes one layer, i.e., business process analysis, function mapping, technical analysis, and human factor analysis. The novelty of the methodology is its holistic nature. Each of the layers contains different and relevant information for security analysis. Moreover, the interrelations between the layers are an interesting source for security analysis. Take for example data flow. There is a process data flow which connects tasks at the business process level by their input and output parameters. In addition, data is exchanged between the process tasks (business process layer) and the invoked applications (system layer), e.g., web services. Based on the methodology, security of process-oriented applications can be investigated. It can be also used for security compliance checking.

In the following, we will use a running example to examine each step of the analysis.

\begin{figure}
	\centering
		\includegraphics[width=\textwidth]{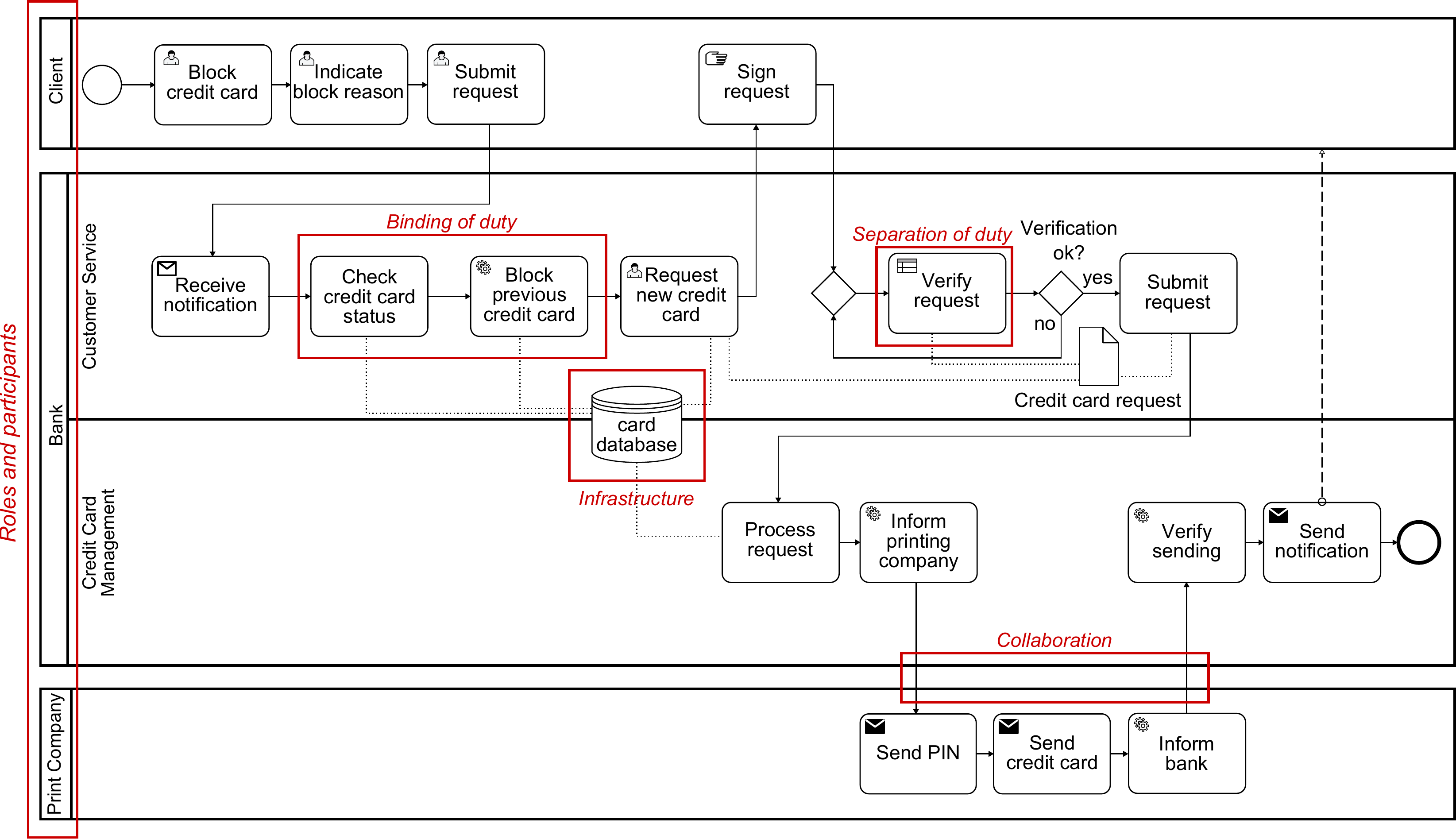}
	\caption{Example~\ref{ex:blockage}: Credit Card Blockage and Renewal}
	\label{fig:requestCreditCard}
\end{figure}

\begin{example}[Credit Card Blockage and Renewal]\label{ex:blockage}
Figure~\ref{fig:requestCreditCard} displays a credit card blockage and renewal request (using BPMN notation, cf. \cite{omg_business_2011}). The process contains several participants: a client that has a credit card, a bank that issues credit cards, and a printing company that prints and sends out the cards. In the example process, a client orders the blockage of a credit card due to card theft or other reasons. Therefore, the client requests the suspension of the previous card and orders a new credit card by signing a request form. The customer service receives the notification of the blocked card, checks the credit status, and blocks the credit card. Furthermore, a binding of duty (BoD) constraint is imposed on the \texttt{Check credit card status} and \texttt{Block previous credit card} tasks, i.e. both tasks have to be performed by the same person. To verify the request, two employees of the customer service have to check and sign the request. This procedure is often called ``four-eyes-principle'' or separation of duty (SoD) constraint (see \cite{botha_separation_2001}). After a successful verification, the request is submitted to the credit card management department that processes the request and informs the printing company. After several working days later the client receives a personal identification number (PIN), a credit card, and an additional notification letter by mail (with several days difference between the letters).
\end{example} 

\subsection{Business Process Analysis}
\label{sec:PreliminaryAnalysis}


Analyzing the business process layer identifies the following information: process tasks as functional building blocks, process participants and their roles, security-related constraints, information about the system infrastructure, and possible business collaborations. Let us illustrate this based on the running example: 
\begin{enumerate}
	\item[(a)] \textit{Identify participants and their roles:} 
Usually, organizational information is (partly) available in business process models. Depending on the notation, it might be represented in a different way. In BPMN as standard notation, pools represented the different involved participants (organizations, partners), specifically, client, bank, and printing company in Figure~\ref{fig:requestCreditCard}. Furthermore, \textsl{roles of process participants} are specified as swimlanes, here the customer service and the credit card management for the bank. If information on participants and roles is not (sufficiently) available, it can be acquired via, e.g., interviews (cf. \cite{leitner_systematic_2014}).
  \item[(b)] \textit{Analyze security-related constraints} such as Separation of Duty (SoD) and Binding of Duty (BoD) which are realized by access control mechanisms that perform authentication and authorization of both users and programs when they access data or use computer resources. In the business process in Figure~\ref{fig:requestCreditCard}, a BoD constraint is defined for tasks \texttt{Check credit card status} and \texttt{Block previous credit card} to ensure that the same person performs both tasks. In addition, the task \texttt{Verify request} contains a business rule that states that two different people have to verify the request. This SoD constraint is imposed to prevent error and fraud (see \cite{botha_separation_2001}).
  \item[(c)] \textit{Denote system infrastructure:} The business process layer partly indicates the \textsl{IT infrastructure} behind the business process. It can be seen from Figure~\ref{fig:requestCreditCard} that a \textit{card database} exists that stores the credit card data. Furthermore, the figure contains automated tasks that connect, for example, the bank with the printing company.
  \item[(d)] \textit{Register business collaborations:} The usage of pools in the figure shows that this business process is a collaboration and therefore might require security measures (e.g., encryption or digital signatures). 
\end{enumerate} 

\subsection{Function Mapping}
\label{sec:FunctionMapping}

The function mapping investigates the functional building blocks (i.e. tasks) of a business process and their incorporation in the system design architecture. 
Note that we are aware that large-scale business processes can contain a vast amount of tasks and the complexity and effort of such analysis would increase quickly. However, in order to minimize the complexity, we recommend to perform the functional mapping and further steps only on tasks that can be classified as security-critical\footnote{In order to determine which tasks are security-critical, for example, the method proposed by \cite{taubenberger2013resolving} can be followed where the decision bases on a discussion between business and security analyst.}. 

\begin{figure}[ht]
	\centering
		\includegraphics[width=\textwidth]{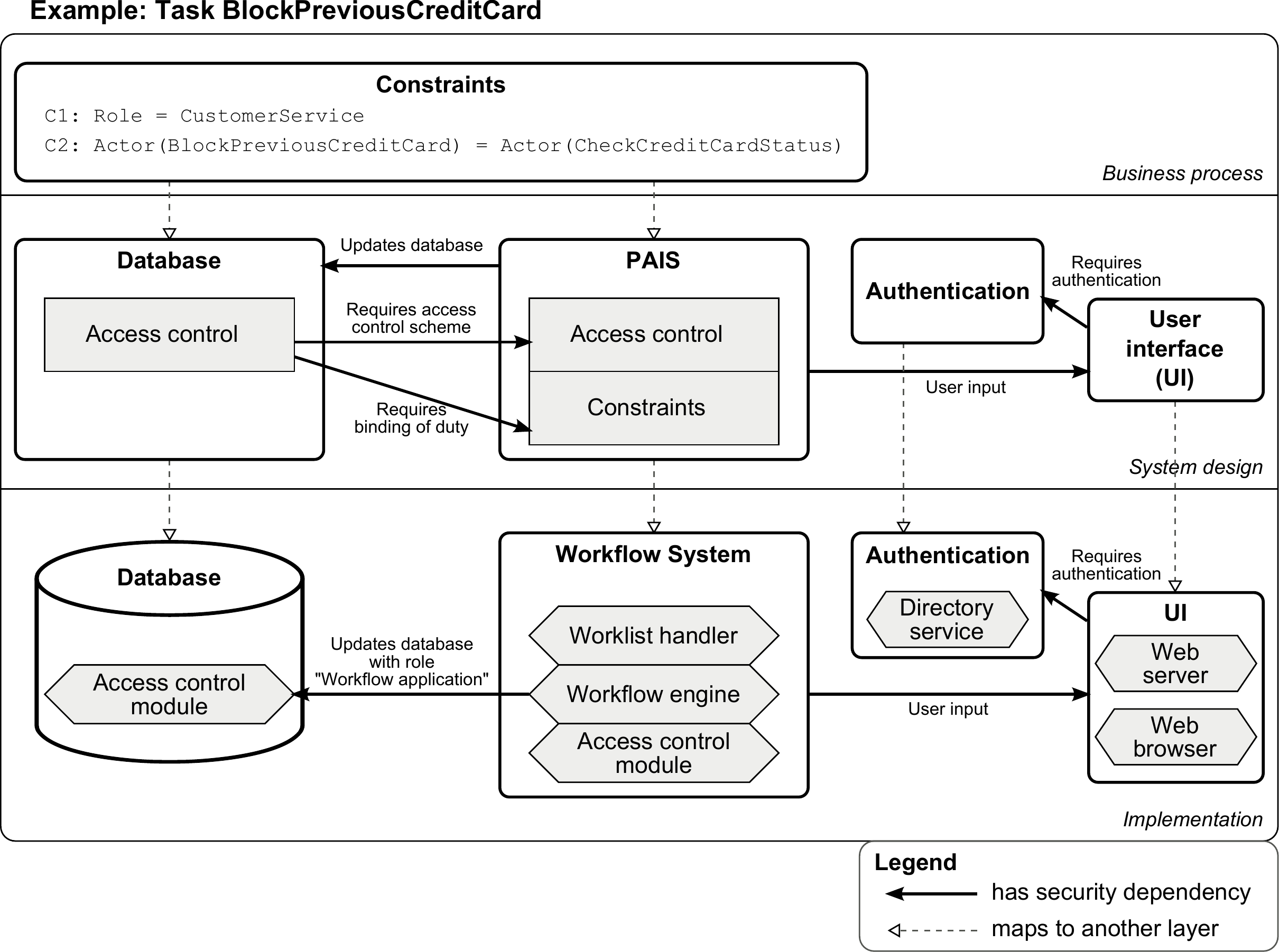}
	\caption{Function Mapping and Technical Analysis of Task \texttt{BlockPreviousCreditCard}}
	\label{fig:BlockPreviousCreditCard}
\end{figure}

Assume that task \texttt{BlockPreviousCreditCard} has been identified as being security-critical. Figure~\ref{fig:BlockPreviousCreditCard} displays its function mapping. The main purpose of the task is to block the credit card and write the required data to a card database. The first layer in Figure~\ref{fig:BlockPreviousCreditCard} displays the constraints at business process layer that can be derived based on the business process model in Figure~\ref{fig:requestCreditCard}. In particular, constraint $C1$ assigns a role to the task and $C2$ defines a BOD constraint (see previous section).

Based on these requirements, we can adopt the function into the system design layer where the systems are outlined and their interconnections. In particular, a database, a PAIS, an authentication module, and a user interface (UI) are identified. The PAIS and database contain modules (e.g., access control) that should align with the constraints defined in the business process layer. 
It can be seen from Figure~\ref{fig:BlockPreviousCreditCard} that depicting the security-aware components and their interactions just for a single task is challenging as it includes different software components that interact with each other. 

\subsection{Technical Analysis}
\label{sec:TechnicalAnalysis}

Based on the system design in the function mapping, the technical analysis is concerned with the deployment and execution of a single task in a business process. 

Continuing with Example~\ref{ex:blockage}, the system infrastructure is explicitly defined in the implementation layer in Figure~\ref{fig:BlockPreviousCreditCard}. 
In particular, the UI and the authentication service enforce a proper authentication and authorization of users. Furthermore, a workflow system interacts with a database in order to store the processed data. Note that in this example, the role accessing the database is not \textit{CustomerService} but is \textit{WorkflowSystem}. Hence, the access control restrictions $C1$ and $C2$ defined in the business process layer are eventually enforced in the workflow system. The user role information is typically not transmitted to the database application.

\begin{figure}[tbh!]
	\centering
		\includegraphics[width=1.0\textwidth]{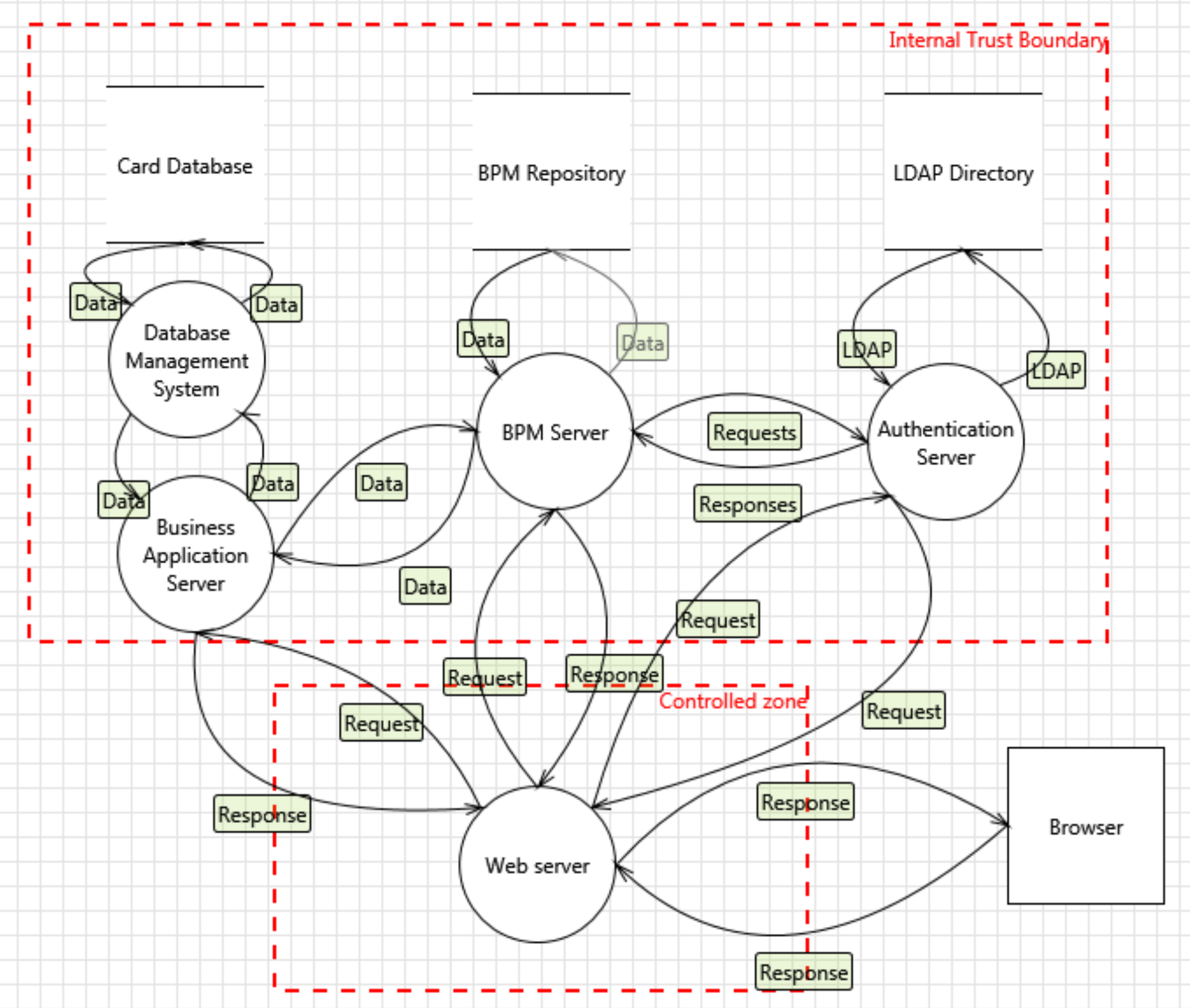}
	\caption{Data flow diagram of system architecture}
	\label{fig:sys-dfd}
\end{figure}

The actual realization of the functions and system depends on the existing enterprise IT environment, available technologies and resources. The result is very likely to be a combination of vendor-specific BPM software and customized software implementations. Figure~\ref{fig:sys-dfd} shows an example of the data flow diagram (DFD) of the system that is often used in threat modeling (cf.~\cite{swiderski2004threat}) for identifying potential IT security threats. All interactions with the user happen on the \textit{Browser}. A \textit{Web server} responds to user requests and data input with constructed web content. The \textit{Authentication Server} handles a user's login requests to the system. The \textit{LDAP Directory} stores the user directory information. The \textit{BPM Server} is where the workflow engine locates which handles the workflow and access control and constraints specified in the BPM rules in the \textit{BPM Repository}. The \textit{Business Application Server} hosts various applications or web services that are invoked by the \textit{BPM server}. The applications on the \textit{Business Application Server} read and write to the \textit{Card database} through the \textit{Data Management System} which provides access control to the \textit{Card Database}, among other functions.  

The access control restrictions $C1$ and $C2$ are thus the function mappings of the requirements to various components and sub-systems. However, this imposes several risks for security: First of all, the constraints defined on the business process layer and drafted on the system design layer are not consistently transferred to the implementation layer. At implementation level, the access control restrictions such as $C1$ and $C2$ are not enforced in the database. The access control restrictions are specified and enforced in the workflow system but the database does not distinguish the updates from the workflow system; they are all made by the role \textit{WorkflowSystem}. Hence, a human-specific verification of the data entries in a database cannot be performed -- only based on the logs of the workflow system. Furthermore, this opens the possibility to access the database given access control restrictions (such as role assignments and constraints) in the workflow system were overcome. 

From an information security point of view, more security risks need to be considered. The system might be distributed geographically and across different trust domains. For example, the main system can be within a \textit{Internal Trust Boundary}, the \textit{Web server} needs to be placed in a \textit{Controlled zone} (e.g. a Demilitary Zone (DMZ)) because it might be subject to various networked attacks originated from the external entities from the internet. The DFD in Figure \ref{fig:sys-dfd} identifies the trust boundaries, data flow, and entry points. In a systematic security analysis, each of the elements need to be considered for security risks. This means even though the access control constraints from the business process layer can be achieved by the functions provided by applications and services such as authentication and workflow engine, any compromise at the system or code level can easily subvert these attempts. Therefore, the functional mapping provide a means for consider business process security and information security at the same time.    

\subsection{Human Factor Analysis}
\label{sec:HumanFactorAnalysis}

Table \ref{tab:humanfactoranalysis} shows example risks and solutions that can be drawn based on the human-oriented cross-layer analysis. Some of the solutions are common security practice. However, other solutions raise interesting new research questions, e.g., consistency checks of access constraints implemented at different layers. 
%
%

\begin{table}[tbh!]
	\centering
	\caption{Human-oriented Security Analysis: Example Risks and Solutions}
		\label{tab:humanfactoranalysis}
		  	\scriptsize
		\begin{tabular}{p{1.8cm}p{0.15cm}p{0.15cm}p{0.15cm}p{0.15cm}p{0.15cm}p{3cm}p{3cm}}
    \toprule
      \textbf{Role} &  \textbf{B}& \textbf{S}& \textbf{I}& \textbf{V} & \textbf{T} & \textbf{Security risk} & \textbf{Solution}\\\hline
    Process designers & $\boxtimes$ & & & $\boxtimes$ & & misconfiguration of access constraints & consistency/compliance checks\\
                      & $\boxtimes$ & & & $\boxtimes$ & & uncontrolled adaption of access constraints/business process& implement change control mechanisms\\
                      &  $\boxtimes$ & & & $\boxtimes$ & & uncontrolled administrative changes & implement change control mechanisms\\
  &   $\boxtimes$ & & & $\boxtimes$ & & miscommunication with stakeholders & regular workshops\\ 
  & & & $\boxtimes$ & $\boxtimes$ & & errors in workflow design & correctness and compliance checks \\\hline
  Developers & & & $\boxtimes$ & $\boxtimes$ & & misconfiguration at application level & follow security best practices and conduct security audit\\
         & & & $\boxtimes$ & $\boxtimes$ & & use default password & change password in configuration\\  
& & & $\boxtimes$ & $\boxtimes$ & & deviate from access constraints as specified at process level & conduct cross-layer consistency checks\\ 
& & & $\boxtimes$ & $\boxtimes$ & & introduce vulnerability in inhouse developed application code & follow security best practices and life cycle\\\hline
  System architects & & $\boxtimes$ & & $\boxtimes$ & & insufficient security controls at different parts of the system & defense in depth  \\
  & & $\boxtimes$ & & $\boxtimes$ & & incorrect definition of trust boundary & conduct thorough security analysis and using reference security architecture  \\
  & & $\boxtimes$ & & $\boxtimes$ & & insufficient protection against (rare) threats  & defense in depth; additional security controls for high value assets \\
   & & $\boxtimes$ & & $\boxtimes$ & & introduce vulnerabilities in the system design enabling user to bypass BPM rules  & conduct cross-layer consistency checks\\\hline
Knowledge workers / Process participants  & &  &$\boxtimes$ & $\boxtimes$ & &  use weak password & security training and technology support for password management   \\
  & &  &$\boxtimes$ &  & $\boxtimes$ &   exploit logical flaws in the workflow & correctness and compliance checks, anomaly detection at BP level  \\
      & &  &$\boxtimes$ &  & $\boxtimes$ &   inject incorrect data into the process data context & data flow checks, implement integrity checks  \\
       & &  &$\boxtimes$ &  & $\boxtimes$ &   inject incorrect data into the database (via the PAIS) & data flow checks, consistent integrity checks between PAIS and database\\\hline
Outsiders & &  &$\boxtimes$ &  & $\boxtimes$ &  launch denial of service attack & design and implement resilience measures such as time out or upper limit for number of instances   \\
 & &  &$\boxtimes$ &  & $\boxtimes$ & social engineering and malware-based attack  &  security training, change management, and security controls  \\
\hline  
\multicolumn{8}{l}{B: Business Process Level; S: System Design Level; I: Implementation Level;} \\	
\multicolumn{8}{l}{ V: Vulnerability, T: Threat} \\				
			
		\end{tabular}
\end{table}

The human factor is analyzed in terms of the roles (cf. Section \ref{sec:HumanLayer}). 
We construct the analysis along the components identified through the functional mapping for task \texttt{BlockPreviousCreditCard} as depicted in Figure~\ref{fig:BlockPreviousCreditCard}. Specifically, a role might introduce a vulnerability or pose a threat for the component (cf. ``negative influences'' of humans on security as outlined in Section \ref{sec:HumanLayer}).

First of all, as we analyze a specific process task, the roles that are responsible for the entire BPM project, i.e., the {\sl BPM group}, the {\sl Business Engineers}, the {\sl Chief process officer}, and the {\sl Process owners} (cf. Section \ref{sec:HumanLayer}) are not considered because they are not directly involved in the business process as shown in Figure~\ref{fig:BlockPreviousCreditCard}. Further on, task \texttt{BlockPreviousCreditCard} neither requires any interaction with {\sl Subjects} nor with {\sl Partners}. Hence, all these roles are not considered within the analysis.

At design time, mostly the technical staff such as system architects or developers are involved because they might introduce errors or make bad design decisions. For example, as software-intensive systems, PAIS often use and integrate off-the-shelf products. Due to complexity and knowledge gap, sometimes software and their security features are not correctly configured, thus becomes vulnerable to malicious attacks. For example, not changing the default authorization value of a SAP service might lead to unrestricted access for all authenticated users (see~\cite{DiCro2011}). A tiny implementation mistake in the OpenSSL library (i.e. the Heartbleed bug) allows an attacker to read sensitive information in a server's memory including the server's private key and other users' passwords (cf.~\cite{heartbleed}).


At runtime, the actual users of PAIS and outsiders can introduce vulnerabilities as well as threats. For example, the access point of knowledge workers or process participants is typically the PAIS. From this point, they can access and execute business process activities. Knowledge workers have an extensive knowledge of the business processes and systems and might seek possibilities to circumvent security controls or exploit knowingly deactivated controls (cf. \cite{epstein_security_2008}). Reconsidering the Heartbleed bug (see \cite{heartbleed}), also a hacker as an outsider can exploit the vulnerability and pose threats to the systems, e.g. by using stolen credentials to log into the system to compromise the whole security controls.
Besides, insecure day-to-day operations by humans can lead to total failure of security protections. Social engineering, which targets humans in daily operations is proved to be quite effective to compromise many high secure systems~\cite{Hadnagy2010}.  

Moreover, the attacks at the implementation can impact other layers such as the system design or the business process layer. For example, if the Heartbleed bug affected the interaction between the database and the workflow system in Example~\ref{ex:blockage}, the PAIS and the business process itself might have been compromised. On the other hand, bad design choices by the technical staff can introduce vulnerabilities on the implementation level that knowledge workers are not aware of. For example, if the system architects do not remove the vulnerability introduced by Heartbleed, the knowledge worker would continue working which can yield data breaches. 
While each single role has a different intersection with the PAIS system, the security implementations can be interrelated. A vulnerability or a threat at one point within one layer might have profound impact or cascading effect on other points at the business process, system design, and implementation layer. Therefore, the presented methodology can bring a big picture of the security issues, which helps us to trace the security interdependency in PAIS at design and runtime and to identify new security issues, in order to find proper solutions.

  \section{Discussion}
\label{sec:Discussion}

This section examines the main findings, impact on research and practice, and the limitations of this paper. 

\subsection{Main Findings}
\label{sec:MainFindings}

This paper provides interdisciplinary research that bridges the gap between BPM/ PAIS and information security research. 
As security is a cross-layer concern, this paper investigates how security is embedded in PAIS architectures. It provides a holistic approach addressing security concerns at the business process, system design, and implementation layer in a PAIS. Moreover, security aspects at each layer are analyzed under consideration of human aspects. Using roles to represent human factors in security analysis gives us a fine-grained grasp of human involvement in various aspect of PAIS. As our analysis shows, it helps to transform the abstract notion of ``human'' into concrete and specific entities such that we can have more precise analysis related to the security implications of specific part of the PAIS architecture. 

The research conducted in this paper shows that security issues are tightly interrelated and the impact of a security attack might be difficult to comprehend without a holistic view on a system. For example, an attack on the implementation to compromise a password can jeopardize the enforcement of constraints defined in the business process layer, or a design error by a system architecture might introduce a vulnerability that creates an opportunity for a disgruntled employee to circumvent the organizationally posed security controls and constraints. Particularly in complex settings with hundreds or thousands of process tasks, a systematic and holistic methodology helps to connect the dots and identify non-obvious security links.  

Security in PAIS is a complex issue. A common approach is to create layers of abstraction, e.g. concentrating issues related to business process, or to software security. Our analysis shows it is also important to consider issues at the interfaces and beyond the abstraction layers.

Overall, the methodology proposed in this paper can be used for 
\begin{itemize}
\item security analysis
\item (security) compliance checks
\item security trainings
\end{itemize}
These usage scenarios will be detailed in the next section. 

\subsection{Impact on Research and Practice}
\label{sec:ImpactOnResearchAndPractice}

A main contribution of this paper is the methodology for a cross-layered and human-oriented security analysis. 
Researchers and practitioners may use this analysis as input for further security evaluations such as risk or threat analysis. For example, the human factor analysis derives a list of internal or external attackers that can be used as a list of ``suspects'' for the IT risk management (e.g., \cite{stoneburner_risk_2002,peltier_information_2005}). The methodology can also serve as input for developing checklists for systematic (security) compliance checks. 
Moreover, this approach can serve as basis for several research directions; the methodology contributes not only to the security engineering of PAIS but also to research on human-centric PAIS (see \cite{kabicher-fuchs_human-centric_2012}). For instance, the training of process participants can raise security awareness, can prevent  misuse or errors, and can increase the knowledge of workers in order to fulfill their job functions successfully. 

Based on the methodology, security-awareness trainings for different target groups such as business partners, developers, or managers can be developed. Furthermore, the result of the analysis can be used for the design and development of security controls. 
In addition, practitioners may use this review for revisiting security measures in PAIS, e.g., by reevaluating existing security measures with regard to human factors. 



\subsection{Limitations}
\label{sec:Limitations}

This paper incorporates roles from the BPM and the information security area. However, we did not include roles from the risk management disciplines (e.g., \cite{stoneburner_risk_2002,peltier_information_2005}) that focus on risk analysis such as employees/ users or chief security officers. One reason is that the classification in this paper centered only on roles from the business process community. It should be noted that some roles from risk management overlap with the roles outlined in Section \ref{sec:HumanLayer}. 

The methodology is evaluated with an application scenario. The scenario should provide an initial guidance how to apply the methodology. 
However, the applicability of the methodology is not restricted by any means: For example, the number of tasks used in the function mapping and technical analysis, the level of detail for the human factor analysis, or the number of vulnerabilities and threats assessed can be adapted accordingly to the scenario.
  \section{Related Work}
\label{sec:RelatedWork}


Humans are an important factor when it comes to information security. Research investigates human users from many perspectives such as trusting, decepting, and hacking humans (cf.~\cite{colwill_human_2009,mitnick_art_2001,okenyi_anatomy_2007}). For example, ways to socially engineer and deceive users in order to acquire information are shown in \cite{mitnick_art_2001}. Furthermore, humans might circumvent security measures (see \cite{kirlappos_learning_2014,bartsch_how_2013}). Other publications such as \cite{sasse_transforming_2001,DBLP:journals/compsec/KraemerCC09} analyze the outcomes of humans errors and use it to provide usable and effective security measures. 
Typically, dependencies can be used in information security to identify and analyze relations such as between assets and attackers (\cite{sawilla_identifying_2008}) or between objectives and patterns (\cite{heyman_using_2008}). In this paper, we analyze dependencies that go beyond the business process in a multilayer model. 

Several approaches tackle risk analysis for business processes. In \cite{ahmed2014securing}, risk patterns for business processes are specified based on BPMN. These patterns partly incorporate the human aspect. The security-risk-oriented pattern 4 (SRP 4), for example, explicitly models the role of an attacker which accesses data in a business process in an unauthorized manner. \cite{ahmed2014securing}, however, is confined to a set of five patterns that foresee possible attackers as explicit roles. By contrast, the approach at hand provides a systematic view on all layers of the system where potential attackers are not known beforehand. 
The approach presented by \cite{taubenberger2013resolving} analyzes a real-world example from the insurance domain with particular focus on the data flow. Hence, it investigates the connections between the tasks rather than the tasks. Such cross-task analyses, in general, can be applied complementary to the function and human-oriented analysis presented in this paper. 
Overall, we found that approaches for security in business processes often concentrate on one aspect such as data flow or security requirements. 


A multilayer architecture in \cite{winter_essential_2006} contains business, process, integration, software, and technology architecture layers. Furthermore, the enterprise architecture is a cross-layer that represents aggregate artifacts and their relationships across all layers. In this paper, security is also seen as a cross-layer issue. 
Furthermore, the secure enterprise business model in \cite{muller_why_2013} is based on a modern enterprise system architecture and contains three layers: First, the \textsl{business layer} specifies business processes and related organizational aspects. Secondly, the \textsl{application layer} specifies the services and data schemas for business process execution. Lastly, the \textsl{infrastructure layer} provides the software and hardware needed to automate the process execution (e.g., PAIS, databases, or operating systems). Furthermore, each layer can be split according to timing: design time, run time, and audit time. I.e. enterprise security and its measures is scattered across all layers and required at all times (design, run, and audit time). While the architecture in \cite{muller_why_2013} centers on the technology and the execution time, our paper provides a methodology that combines technical and human aspects. We do not consider audit time in this paper since the respective analysis examines the security impacts on the systems during design- and runtime. To our best knowledge we are not aware of any other literature that investigates the human factor in business process security. 

Furthermore, a case study on modeling secure transactions is shown in \cite{giorgini_requirement_2003}. In particular, the authors use Secure Tropos (see \cite{mouratidis_secure_2007}) to model dependencies between merchants that offer services, banks that requesting the service and cardholders that own the data. In this paper, we analyze security based on a payment card industry use case. However, we do not define specific requirements nor provide a methodology to define security-related artifacts. In fact, we use this as example to display that security is a issue across all layers. In addition, we analyze the dependencies between the levels in a multilayer architecture model. 
Furthermore in \cite{lowis_vulnerability_2011}, the attentive listener (ATLIST) vulnerability analysis is provided. This method was mainly developed for service-oriented architectures that use a plethora of web technologies and SOA-specific standards and targets SOA-based business processes. Our methodology can be used to determine prerequisite artifacts for the ATLIST analysis.

  \section{Conclusion}
\label{sec:Conclusion}

Security is as good as its weakest link -- the humans -- in everyday business processes. So far, research has neglected to analyze the human factor when it comes to vulnerabilities and threats in PAIS. 
This paper proposes a cross-layer and human-oriented methodology for PAIS. 
The methodology examines business functionality beyond the business process layer and investigates technical and human aspects that are essential for a thorough examination. With an application scenario, we show that even the security analysis for a single business function results into an assessment of a multilateral interaction between multiple application components. Overall, the proposed methodology contributes to the security engineering of PAIS which can be used as a basis to conduct further security evaluations, systematic compliance checking, and target-group oriented security trainings. As a next step, the application of the analysis will be applied to selected real-world business process scenarios. 


%
%

\bibliography{BookChapter,ref2}

\begin{thebibliography}{10}

\bibitem{ahmed2014securing}
Naved Ahmed and Raimundas Matulevi{\v{c}}ius.
\newblock Securing business processes using security risk-oriented patterns.
\newblock {\em Computer Standards \& Interfaces}, 36(4):723--733, 2014.

\bibitem{bartsch_how_2013}
Steffen Bartsch and M.~Angela Sasse.
\newblock How users bypass access control - and why: The impact of
  authorization problems on individuals and the organization.
\newblock In {\em {ECIS} 2013 Completed Research}, 2013.

\bibitem{bohr2013business}
Frank B{\"o}hr, Linh~Thao Ly, and G{\"u}nter M{\"u}ller.
\newblock Business process security analysis--design time, run time, audit
  time.
\newblock {\em it--Information Technology it--Information Technology},
  55(6):217--224, 2013.

\bibitem{botha_separation_2001}
Reinhardt~A. Botha and Jan H.~P. Eloff.
\newblock Separation of duties for access control enforcement in workflow
  environments.
\newblock {\em {IBM} Systems Journal}, 40(3):666--682, 2001.

\bibitem{briney_security_2000}
Andy Briney.
\newblock Security focused.
\newblock {\em Information Security}, pages 40--68, 2000.

\bibitem{carminati_web_2005}
B.~Carminati, E.~Ferrari, and {P.C.}~K. Hung.
\newblock Web service composition: A security perspective.
\newblock In {\em Workshop on Challenges in Web Information Retrieval and
  Integration}, pages 248--253. {IEEE}, 2005.

\bibitem{Casey:2010:TAN:1852666.1852728}
Timothy Casey, Patrick Koeberl, and Claire Vishik.
\newblock Threat agents: A necessary component of threat analysis.
\newblock In {\em Workshop on Cyber Security and Information Intelligence
  Research}, pages 56:1--56:4. ACM, 2010.

\bibitem{heartbleed}
{Codenomicon Ltd.}
\newblock The heartbleed bug.
\newblock Accessed: 2015-05-04.

\bibitem{colwill_human_2009}
Carl Colwill.
\newblock Human factors in information security: The insider threat – who can
  you trust these days?
\newblock {\em Information Security Technical Report}, 14(4):186--196, 2009.

\bibitem{dhillon2001current}
Gurpreet Dhillon and James Backhouse.
\newblock Current directions in is security research: towards
  socio-organizational perspectives.
\newblock {\em Information Systems Journal}, 11(2):127--153, 2001.

\bibitem{dumas_fundamentals_2013}
M~Dumas, M.~La~Rosa, J.~Mendling, and H.~A. Reijers.
\newblock {\em Fundamentals of Business Process Management}.
\newblock Springer, 2013.

\bibitem{epstein_security_2008}
Jeremy Epstein.
\newblock Security lessons learned from soci\'{e}t\'{e} g\'{e}n\'{e}rale.
\newblock {\em {IEEE} Security Privacy}, 6(3):80--82, 2008.

\bibitem{fuchs_roles_2011}
L.~Fuchs, G.~Pernul, and R.~Sandhu.
\newblock Roles in information security - a survey and classification of the
  research area.
\newblock {\em Computers \& Security}, 30(8):748--769, 2011.

\bibitem{giorgini_requirement_2003}
Paolo Giorgini, Fabio Massacci, and John Mylopoulos.
\newblock Requirement engineering meets security: A case study on modelling
  secure electronic transactions by {VISA} and mastercard.
\newblock In {\em Conceptual Modeling}, pages 263--276. Springer, 2003.

\bibitem{granger_social_2001}
Sarah Granger.
\newblock Social engineering fundamentals, part i: hacker tactics, 2001.
\newblock Accessed: 2014-04-07.

\bibitem{Hadnagy2010}
Christopher Hadnagy.
\newblock {\em Social Engineering: The Art of Human Hacking}.
\newblock Wiley, 2010.

\bibitem{hafner_seaas-reference_2009}
M.~Hafner, M.~Memon, and R.~Breu.
\newblock {SeAAS-A} reference architecture for security services in {SOA}.
\newblock {\em Journal of Universal Computer Science}, 15(15):2916--2936, 2009.

\bibitem{heyman_using_2008}
T.~Heyman, R.~Scandariato, C.~Huygens, and W.~Joosen.
\newblock Using security patterns to combine security metrics.
\newblock In {\em Availability, Reliability and Security}, pages 1156--1163,
  2008.

\bibitem{jensen_soa_2007}
Meiko Jensen, Nils Gruschka, Ralph Herkenh\"oner, and Norbert Luttenberger.
\newblock {SOA} and web services: New technologies, new standards - new
  attacks.
\newblock In {\em European Conference on Web Services}, pages 35--44, 2007.

\bibitem{kabicher-fuchs_human-centric_2012}
Sonja Kabicher-Fuchs, Stefanie Rinderle-Ma, Jan Recker, Marta Indulska,
  Francois Charoy, Rob Christiaanse, Reinhold Dunkl, Gregor Grambow, Jens Kolb,
  Henrik Leopold, and Jan Mendling.
\newblock Human-centric process-aware information systems ({HC-PAIS)}.
\newblock Technical report, arxiv.org, 2012.

\bibitem{kavantzas_web_2005}
N.~Kavantzas, David Burdett, Gregory Ritzinger, Tony Fletcher, Yves Lafon, and
  Charlton Barreto.
\newblock Web services choreography description language version 1.0.
\newblock Technical report, 2005.
\newblock {http://www.w3.org/TR/ws-cdl-10/}.

\bibitem{kirlappos_learning_2014}
I~Kirlappos, S~Parkin, and M.~Angela Sasse.
\newblock Learning from "shadow security": Why understanding non-compliance
  provides the basis for effective security.
\newblock In {\em Workshop on Usable Security}, 2014.

\bibitem{DBLP:journals/compsec/KraemerCC09}
Sara Kraemer, Pascale Carayon, and John Clem.
\newblock Human and organizational factors in computer and information
  security: Pathways to vulnerabilities.
\newblock {\em Computers {\&} Security}, 28(7):509--520, 2009.

\bibitem{leitner_systematic_2014}
Maria Leitner and Stefanie Rinderle-Ma.
\newblock A systematic review on security in process-aware information systems
  – constitution, challenges, and future directions.
\newblock {\em Information and Software Technology}, 56(3):273--293, 2014.

\bibitem{lowis_vulnerability_2011}
Lutz Lowis and Rafael Accorsi.
\newblock Vulnerability analysis in {SOA-Based} business processes.
\newblock {\em {IEEE} Transactions on Services Computing}, 4(3):230--242, 2011.

\bibitem{mitnick_art_2001}
Kevin~D. Mitnick and William~L. Simon.
\newblock {\em The Art of Deception: Controlling the Human Element of
  Security}.
\newblock John Wiley \& Sons, September 2001.

\bibitem{mouratidis_secure_2007}
{Haralambos} {Mouratidis} and {Paolo} {Giorgini}.
\newblock Secure tropos: A security-oriented extension of the tropos
  methodology.
\newblock {\em International Journal of Software Engineering and Knowledge
  Engineering}, 17(02):285--309, 2007.

\bibitem{mouratidis_security_2007}
Haralambos Mouratidis and Paolo Giorgini.
\newblock Security attack testing ({SAT)—testing} the security of information
  systems at design time.
\newblock {\em Information Systems}, 32(8):1166--1183, 2007.

\bibitem{muller_why_2013}
G\"unter M\"uller and Rafael Accorsi.
\newblock Why are business processes not secure?
\newblock In {\em Number Theory and Cryptography}. Springer, 2013.

\bibitem{DiCro2011}
Mariano Nu\~nez Di~Croce.
\newblock Attacks to {SAP} web applications.
\newblock BlackHat DC 2011 Briefings, 2011.

\bibitem{okenyi_anatomy_2007}
Peter~O. Okenyi and Thomas~J. Owens.
\newblock On the anatomy of human hacking.
\newblock {\em Information Systems Security}, 16(6):302--314, 2007.

\bibitem{omg_business_2011}
{OMG}.
\newblock Business process model and notation ({BPMN)} version 2.0.
\newblock {OMG} Document formal/2011-01-03, Object Management Group, January
  2011.

\bibitem{peltier_information_2005}
Thomas~R. Peltier.
\newblock {\em Information Security Risk Analysis, Second Edition}.
\newblock {CRC} Press, April 2005.

\bibitem{rodriguez_bpmn_2007}
Alfonso Rodriguez, Eduardo Fernandez-Medina, and Mario Piattini.
\newblock A {BPMN} extension for the modeling of security requirements in
  business processes.
\newblock {\em {IEICE} {TRANSACTIONS} on Information and Systems},
  E90-D(4):745--752, April 2007.

\bibitem{sasse_transforming_2001}
M.~A. Sasse, S.~Brostoff, and D.~Weirich.
\newblock Transforming the 'weakest link' - a {Human/Computer} interaction
  approach to usable and effective security.
\newblock {\em {BT} Technology Journal}, 19(3):122--131, 2001.

\bibitem{sawilla_identifying_2008}
Reginald~E. Sawilla and Xinming Ou.
\newblock Identifying critical attack assets in dependency attack graphs.
\newblock In {\em Computer Security}, pages 18--34. Springer, 2008.

\bibitem{stoneburner_risk_2002}
Gary Stoneburner, Alice Goguen, and Alexis Feringa.
\newblock Risk management guide for information technology systems.
\newblock {NIST} special publication 800-30, National Institute of Standards
  and Technology ({NIST)}, July 2002.

\bibitem{swiderski2004threat}
Frank Swiderski and Window Snyder.
\newblock {\em Threat modeling}.
\newblock Microsoft Press, 2004.

\bibitem{taubenberger2013resolving}
Stefan Taubenberger, Jan J{\"u}rjens, Yijun Yu, and Bashar Nuseibeh.
\newblock Resolving vulnerability identification errors using security
  requirements on business process models.
\newblock {\em Information Management \& Computer Security}, 21(3):202--223,
  2013.

\bibitem{verizon_2014_2014}
Verizon.
\newblock 2014 data breach investigations report.
\newblock Technical report, 2014.

\bibitem{vorobiev_security_2006}
A.~Vorobiev and Jun Han.
\newblock Security attack ontology for web services.
\newblock In {\em Semantics, Knowledge and Grid, 2006}, pages 42--48, 2006.

\bibitem{walton_balancing_2006}
Richard Walton and Walton-Mackenzie Limited.
\newblock Balancing the insider and outsider threat.
\newblock {\em Computer Fraud \& Security}, (11):8--11, 2006.

\bibitem{weske_business_2007}
M.~Weske.
\newblock {\em Business Process Management: Concepts, Languages,
  Architectures}.
\newblock Springer, 2007.

\bibitem{winter_essential_2006}
R.~Winter and R.~Fischer.
\newblock Essential layers, artifacts, and dependencies of enterprise
  architecture.
\newblock In {\em Enterprise Distributed Object Computing Conference Workshops
  (EDOCW)}, pages 30--30. IEEE, 2006.

\bibitem{wolter_modelling_2008}
C.~Wolter, M.~Menzel, and C.~Meinel.
\newblock Modelling security goals in business processes.
\newblock In {\em Modellierung}, pages 197--212. GI, 2008.

\end{thebibliography}

\end{document}